\documentclass[10pt,conference]{IEEEtran}
\IEEEoverridecommandlockouts
\usepackage{fontawesome}
\newcommand{\newcheckmark}{\textrm{\faCheck}}
\newcommand{\newcrossmark}{\textrm{\faTimes}}
\usepackage{amsmath}
\usepackage{bm}
\AtBeginDocument{%
  \providecommand\BibTeX{{%
    Bib\TeX}}}
\usepackage{cite}
\usepackage{amsmath,amssymb,amsfonts}
\usepackage{algorithmic}
\usepackage{graphicx}
\usepackage{textcomp}
\usepackage{xcolor}
\usepackage{tabularx}
\def\BibTeX{{\rm B\kern-.05em{\sc i\kern-.025em b}\kern-.08em
    T\kern-.1667em\lower.7ex\hbox{E}\kern-.125emX}}
\usepackage{tcolorbox}
\usepackage{multirow}
\usepackage[normalem]{ulem}
\useunder{\uline}{\ul}{}

\newcommand{\realsearchgoal}{\textbf{\textit{ the goal of this study is to aid practitioners in producing more secure software products and make informed decisions on the security practices of candidate dependencies by depicting the current security practices and gaps across ecosystems via an empirical study of the OpenSSF Scorecard project.
}}}

\begin{document}
\author{\IEEEauthorblockN{Nusrat Zahan, Parth Kanakiya, Brian Hambleton, Shohanuzzaman Shohan, Laurie Williams}
\IEEEauthorblockA{Email: [nzahan, pkanaki, bthamble, sshohan, lawilli3]@ncsu.edu\\
North Carolina State University, Raleigh, USA
}}

\title {OpenSSF Scorecard:  On the Path Toward Ecosystem-wide Automated Security Metrics}

\maketitle
\begin{abstract}
The OpenSSF Scorecard project is an automated tool to monitor the security health of open-source software.   This study evaluates the applicability of the Scorecard tool and compares the security practices and gaps in the npm and PyPI ecosystems. 

\end{abstract}

\section{\textbf{Introduction}}
The 2022 annual report from Sonatype shows an average 742\% annual increase in software supply chain (SSC) attacks over the past three years~\cite{sonatype_2022}. %
Therefore, practitioners are increasingly concerned with whether their projects' open-source components are secure. 

Though standards, such as the NIST Secure Software Development Framework (SSDF)~\cite{souppaya2022secure} and OWASP Software Component Verification Standard (SCVS)~\cite{scvs}, provide exhaustive lists of security practices, a lack of consensus is observed regarding the implementation, validation, and verification of these practices towards a unified and consistent baseline measurement. Research is being conducted on the development of different security metrics. However,  establishing a pipeline to measure security is not straightforward since it involves exploring various sources of information, including source code repositories, vulnerability tracking systems, continuous integration/ continuous deployment (CI/CD) pipelines, license(s) validity, package release history, and other metrics to develop standards for adoption. Additional challenges arise during the security assessment of packages in a software supply chain, particularly when the packages come from different sources and have different functionality. 
Practitioners want to make informed decisions about whether or not packages meet security standards based on evidence. Also, practitioners desire to monitor the ``health" of open-source-software (OSS) to identify and manage any future risks of the software supply chain attacks. 
Therefore, practitioners are more interested than ever in identifying healthy open-source components and determining the security practices compared to other components within the ecosystem. Towards this end, \realsearchgoal

The OpenSSF Scorecard project~\cite{Scorecard} is an automated tool to monitor the security health of the OSS supply chain. The primary goal of this project is to auto-generate a ``security score” for OSS projects, using a list of 18 security metrics that can be used to assess the security health of potential dependencies. While projects like Scorecard exist to perform heuristic-based checks of a package's security practices to aid dependency selection, little research has been done to understand the viability of using Scorecard security metrics to identify existing security gaps and practices in an entire ecosystem in addition to the individual packages.  Observing the pattern of these security metrics across one or more ecosystems can assist practitioners in determining how their packages fit into the ecosystem and what they can do to improve security. Practitioners can also benefit by knowing whether a specific security metric is effective within that ecosystem. In this work, we studied the Scorecard tool to evaluate the tool's applicability and analyze what security practice patterns are observed in both ecosystems.

\section{\textbf{OpenSSF Scorecard}} \label{Scorecard}
The Open Source Security Foundation (OpenSSF), sponsored by the Linux Foundation, is a cross-industry collaboration with a mission to improve OSS's security. %
OpenSSF launched the Scorecard project~\cite{Scorecard} in November 2020 to provide an automated security tool that gives a ``security score" for OSS and reduces the manual effort required to analyze a package's security. These results are made available via a BigQuery public dataset and the Open Source Insight (OSI) site. Additionally, practitioners can execute Scorecard on a specific GitHub repository to evaluate the security practices of that repository. 

At the time of the study, the Scorecard contained 18 security practice metrics and assigned an ordinal score between 0 to 10 to each. 
Each metric has one of four risk levels: \textbf{``Critical"} risk-weight 10; \textbf{``High”} risk-weight 7.5; \textbf{``Medium”} risk-weight 5; and \textbf{``Low”} risk-weight 2.5. An aggregate confidence score is also provided, which is a weighted average of the individual metric scores weighted by risk. Table \ref{tab:scorecards} provides information on the 18 Scorecard metrics.

\begin{table*}  \renewcommand{\arraystretch}{1.2}
\caption{\centering{Scorecard Security Metrics and the mapping to the SSDF Framework (ranked from critical to low risk)}} \label{tab:scorecards} 
\centering \footnotesize
\begin{tabular}{| p{60pt} || p{290pt} ||p{60pt}| } \hline
Metrics Name (Risk Label) & \centering{Security Metrics Description} & Mapping to SSDF Practices  \\
\hline\hline
\textbf{Dangerous-Workflow} (Critical) & Indicates if there are dangerous patterns in the package's \texttt{GitHub workflows} due to misconfigured GitHub Actions. %
A list of event context data, such as GitHub issues or pull requests, can be controlled by users and, if exploited, may lead to malicious injection.  & \\\hline

\textbf{Vulnerabilities} (High) & Indicates the presence of unfixed vulnerabilities of a package in the Open Source Vulnerabilities (OSV)~\cite{osv_adv} database.  &\textbf{PW.4, RV.1}\\\hline

\textbf{Binary-Artifacts} (High) & Indicates the presence of executable (binary) artifacts in the repository. Since binary artifacts cannot be reviewed, it is possible to maliciously subvert the executable. & \\\hline

\textbf{Token-Permissions} (High) & Indicates whether the package's automated workflow tokens are set to read-only. This is important because attackers might inject malicious code into the project using a compromised token with write access. If the permission's definitions in each workflow's \texttt{yaml} file are set as read-only at the top level, and the required write permissions are declared at the run-level, the project gets the highest score.  &\textbf{PO.5, PS.1}\\\hline

\textbf{Code-Review} (High) & Indicates if the practitioners conducts code reviews prior to merging a PR. %
The first step of the check is to see if Branch-Protection is activated with at least one required reviewer. If the step fails, the check looks to see if the last 30 commits are Prow, Gerrit, Github-approved reviews or if the merger differs from the committer. & \textbf{PW.7, RV.1}\\\hline

\textbf{Maintained} (High) & Indicates if the package is actively maintained and obtains the score based on activities on commits and issues from collaborators, members, or project owners. For example, if a project has at least one commit per week for the preceding 90 days for the latest 30 commits and issues, it will receive the highest score. Inactive projects run the risk of having unpatched code and insecure dependencies. & \textbf{PW.4}\\\hline

\textbf{Branch-Protection} (High) & Indicates whether GitHub's branch protection settings have been applied to a package's branches. This check enables maintainers to set guidelines to enforce specific workflows, such as requiring reviews or passing particular status checks before acceptance into the main branch. The check is scored on a five-tiered scale. Each tier has multiple checks and must be fully satisfied to gain points at the next tier. & \textbf{PS.1}\\\hline

\textbf{Dependency-Update-Tool} (High) & Indicates whether the repository has enabled dependabot or renovatebot  dependency update tool to automate the process of updating outdated dependencies by opening a pull request. Out-of-date dependencies are prone to attacks. & \textbf{PO.3, PW.4} \\\hline

\textbf{Signed-Releases} (High) & Indicates whether the project signed the release artifacts in GitHub by looking for the following filenames in the project's last five releases: \texttt{*.minisig, *.asc (pgp), *.sig, *.sign}. Signed-Releases attest to the provenance of the artifact. & \textbf{PS.1, PS.2, PS.3}\\\hline

\textbf{Pinned-Dependencies} (Medium) &Indicates unpinned dependencies in \texttt{Dockerfiles, shell scripts}, and \texttt{GitHub workflows} to verify the project's locked dependencies. Unpinned-Dependency allows auto-updating a dependency to a new version without reviewing the differences between the two versions,  which may include an insecure component.  & \\\hline

\textbf{Security-Policy} (Medium) & Report on a file entitled \texttt{SECURITY.md}(case-insensitive) in directories like the top-level or the \texttt{.github} of a repository to see if the package has published a security policy. Users can learn what constitutes a vulnerability and how to report it securely via a security policy. & \textbf{RV.1}\\\hline

\textbf{Packaging} (Medium) & Indicates language-specific GitHub Actions that upload the package to a related hub and determines if the package is published by GitHub packaging workflows. Packaging makes it easy for users to receive security patches as updates. & \\\hline

\textbf{Fuzzing} (Medium) & Indicates if the project uses fuzzing by checking the repository name in the OSS-Fuzz project list. Fuzzing is important to detect exploitable vulnerabilities.  & \textbf{PW.8}\\\hline

\textbf{Static Application Security Testing (SAST)} (Medium) & Indicates if the project uses SAST. These tools can prevent bugs from being inadvertently introduced in the codebase. The metric look for known Github apps such as CodeQL, LGTM, and SonarCloud in the recent merged PRs, or the use of ``\texttt{GitHub/codeql-action}" in a GitHub workflow. & \textbf{PW.7, PW.8}\\\hline

\textbf{License} (Low)& Indicates if the project has published a license by looking for any combination of the following names and extensions in the top-level directory: \texttt{LICENSE, LICENCE, COPYING, COPYRIGHT} and \texttt{.html,.txt,.md}. Scorecard can also detect these files in the \texttt{LICENSES} directory. The lack of a license will hinder any security review and create a legal risk for potential users. & \\\hline

\textbf{CII-Best-Practices} (Low) & Indicates whether the package has a CII Best Practices Badge, which certifies that it follows a set of security-oriented best practices such as vulnerability reporting policy, automatic process to rebuild the software, SAST, and so on. & \textbf{PS.1, PS.2 RV.1, PW.5, PW.8} \\\hline

\textbf{CI-Tests} (Low) & Indicates if the project runs tests before PRs are merged by looking for a set of CI-system names in GitHub \texttt{CheckRuns} and \texttt{Statuses} in recent 30 commits. CI-Tests enable developers to identify problems early in the pipeline. & \textbf{RV.1}\\\hline

\textbf{Contributors} (Low) & Indicates if the project has contributors from multiple organizations by looking at the company field on the GitHub user profile to identify trusted code reviewers. The project must have had contributors from at least three organizations in the last 30 commits to receive the highest score.  & \\\hline
\end{tabular}
\end{table*}

\section{\textbf{Security frameworks}}
\textbf{Guidelines and Standards: }
The OWASP \textbf{Software Component Verification Standard (SCVS)}~\cite{scvs} is a framework to develop a common set of activities, controls, and security practices that can help in identifying and reducing risk in a software supply chain. There are $6$ control families that contain $87$ controls for different aspects of security verification or processes. The SCVS has three verification levels, where higher levels include additional controls.

In response to Section 4 of the President’s Executive Order (EO) on “Improving the Nation’s Cybersecurity (14028)”~\cite{EO_2021}, the U.S. National Institute of Standards and Technology (NIST) updated the \textbf{Secure Software Development Framework (SSDF)}~\cite{souppaya2022secure}. The framework comprises four groups containing high-level security practices and tasks based on established secure software development models. Each group has a number of practices, which are further split into different tasks. These four groups are-
\begin{itemize}
    \item \textbf{Prepare the Organization (PO)}: Practices-5, Tasks-13 
    \item \textbf{Protect the Software (PS)}: Practices-3, Tasks-4 
    \item \textbf{Produce Well-Secured Software (PW)}: Practices-9, Tasks-16 
    \item \textbf{Respond to Vulnerabilities (RV)}: Practices-3, Tasks-9).
\end{itemize}

Automation is essential for implementing security practices at scale.
The Scorecard tool allows us to automate the measurement of security practice metrics at scale. We investigated whether the 18 security practices defined by Scorecard~\cite{Scorecard} complement the Executive Order (EO) and SSDF framework as part of secure SDLC practices for organizations. %
To that end, two authors individually mapped each metric to SSDF practices and compared the findings. We found that out of the 18 Scorecard security metrics, 13 can be mapped to the SSDF framework's practices. Table \ref{tab:scorecards} showed the mapping between each Scorecard metric and SSDF practices. Note that each SSDF practice consists of a number of tasks, hence, a practice can be linked to more than one Scorecard metrics.

\section{\textbf{Methods}}
This section discusses the data sourcing and generation process of this study. We compiled a package list and relevant metadata from the npm and PyPI ecosystems to collect the security score for those packages from Scorecard tool.

\subsection{\textbf{Ecosystem package metadata}} \label{metadata}
\medskip

\textbf{Package name:} %
To begin, we collected a list of all package names available in both ecosystems. We sourced the list of npm packages names (1,494,105) from study~\cite{zahan2022weak} and the list of PyPI package names (365,450) was collected using PyPI API~\cite{PyPI_API} %
in April 2022. 

\textbf{Dependents data: } The number of dependents reflects the importance of a project by quantifying how many other projects use it. We collected dependent information from the  OSI API~\cite{OSI}, a Google-developed and hosted tool. %
In this work, we collected dependent information to prioritize the packages list for manual review. %

\subsection{\textbf{OpenSSF Scorecard score}} 
The Scorecard tool only runs on source code hosted by GitHub. Hence, to obtain the Scorecard scores for a given package, the first step was to map the package to its respective source code location. To retrieve the source code location for both ecosystems, we use the OSI API~\cite{OSI}. We collected unique GitHub repositories of 767,389 npm and 191,158 PyPI packages. The package-to-repository mapping is not always a 1:1 match. Multiple packages can be found in a single repository. In total, we collected 947,936 npm packages with 767,389 unique GitHub repositories and 211,088 PyPI packages with 191,158 unique GitHub repositories. 

Then, Scorecard runs a weekly scan of %
open-source packages to generate the security score of those packages. %
However, we could not directly utilize this data for both ecosystems because, at the time of this study, Scorecard scores were only generated on 760k of 947K npm and 10K of 211K PyPI packages. Therefore, we submitted a pull request to the Scorecard repository, adding the GitHub repositories of missing packages to collect the scores  %
from both ecosystems. The weekly Scorecard scan was able to run on those GitHub repositories after the Scorecard team successfully merged the PR. 

Out of the 947,936 npm packages and 211,088 PyPI packages, we collected the generated score of 832,422 npm packages and 191,483 PyPI packages. We reviewed 50 randomly-chosen packages where the Scorecard failed to generate scores and found that we did not have access to those GitHub repositories. We collected the Scorecard score on May 09, 2022. 

For each package, we could obtain 15 out of 18 Scorecard security metrics and their aggregate score, with the missing 3 metrics being the CI-Test, SAST, and CONTRIBUTOR metrics. The Scorecard team took out these three metrics to scale the weekly job since computing these metrics is API intensive, and GitHub rate limiting can be a bottleneck for the weekly run. As a result, we could not collect data for these three metrics. %

\begin{table*}
  \caption{npm and PyPI Ecosystems Security Practices measured by Scorecard Tool}
  \label{tab:final_data}
  \includegraphics[width=\linewidth]{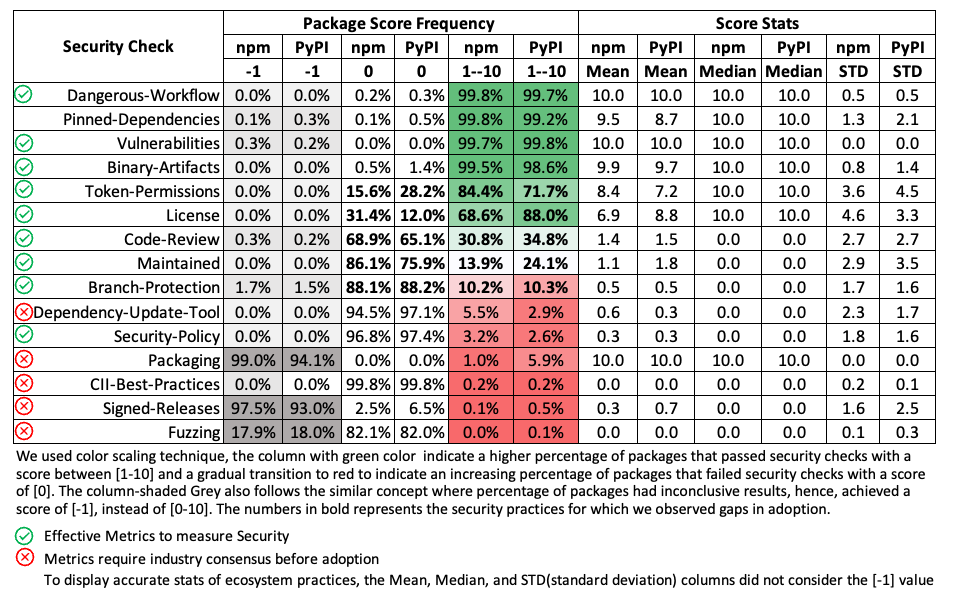}
\end{table*}

\subsection{\textbf{Scorecard metrics evaluation}}
To \textbf{evaluate} the \textbf{Scorecard tool} metrics and learn why a metric passed or failed, we manually reviewed 25 sample GitHub repositories from each ecosystem for each practice. We ranked each metric by the highest number of dependents and selected the top 25 packages. One author reviewed 50 GitHub repositories (25 from each ecosystem) %
totaling 750 repositories for the 15 metrics. A second reviewer then verified the findings by selecting 100 repositories at random. We used the Cohen Kappa statistic to test the inter-rater reliability and achieved a score of 0.96. We resolved our disagreement after discussing our findings, and the first reviewer cross-reviewed other repositories to make changes if required. Then, we needed to examine more packages to understand a given score for Vulnerabilities and Code-Review metrics. We again chose further packages by highest dependent order.

\subsection{\textbf{Ecosystems security practices evaluation}} We observe each ecosystem's security practices and patterns by comparing the Scorecard security metrics scores in three categories: [-1, 0, 1-10]. For each of the 15 security practices, the frequency of packages is measured using these categories ([-1, 0, 1-10]), as shown in Table \ref{tab:final_data}. %

The notation ``$-1$'' denotes the percentage of packages achieving a score of -1 and indicates that Scorecard could not get conclusive evidence of the practice being implemented, or perhaps an internal error occurred due to a runtime error in Scorecard. The notation ``$0$'' denotes the percentage of packages achieving a score of 0 and means that the tool reports indicate the practice was conclusively determined not to be implemented. Since a value of 0 will affect the package's aggregate score, Scorecard assigned a value of -1 to avoid the penalty of failing a metric and also to distinguish between conclusive and inconclusive outcomes. Seven of the 15 security metrics had packages with a score of $-1$.  %

The notation ``$1-10$" denotes the percentage of packages achieving scores ranging from 1 to 10. In table \ref{tab:final_data}, the [1-10] columns display the frequency of  npm and PyPI package scores in descending order. Additionally, the mean, median, and standard deviation (STD) are measured to understand an ecosystem's central tendency and spread of score distribution. 

\section{\textbf{Results}}
This section discusses the finding of our study.  While evaluating Scorecard metrics and ecosystem security practices, we looked into whether Scorecard metrics-
\begin{itemize}
\item Are effective for npm and PyPI ecosystems. %
\item Require improvement of Scorecard tool.
\item Require industry consensus to identify standard practices%
\item Reported lack of adoption of security practices in npm and PyPI ecosystems
\end{itemize}

In Table \ref{tab:final_data}, a higher \% (green cell) in ``$1-10$'' column shows that more than 50\% packages implemented the practices. A lower \% (red cell) indicates more than 50\% of packages failed the practice and received a score of 0 or -1.
Then the green checkmark (\newcheckmark) beside nine metrics represents the metrics that can be used to measure security by Scorecard. The red crossmark (\newcrossmark) besides five metrics indicates that the guideline proposed by Scorecard was not adopted in practice, requires industry consensus due to metrics inheritance reliance on other systems and more than 90\% packages in both ecosystems scored either 0 or -1.  We discuss each security metric for both ecosystems and the frequency statistics in the following subsection. We also highlight the efficacy of Scorecard security guidelines.

\subsubsection{Dangerous-Workflow}%

This metric indicates the following two patterns in workflows: untrusted code checkout; and script injection with untrusted context variables. More than 99\% packages passed the metric. However, we observed 1,938 (0.2\%) npm packages and 508 (0.3\%) PyPI packages where Scorecard found vulnerable code patterns. Out of the 50 repositories used for manual analysis, we had 8 packages with $-1$, all of which were the outcome of internal errors, and 11 packages with vulnerable code patterns in workflows, hence, scored $0$. Among them, 3 npm packages had untrusted code checkout patterns, and 5 PyPI and 2 npm packages had warnings about script injection. %
At the end of this section, we provide a case study explaining how an attacker can exploit such patterns in workflows. 

Additionally, we observe that Scorecard reports a score of 10 for Dangerous-Workflow metrics in empty repositories because the repositories did not have any GitHub workflows let alone dangerous patterns, and the tool lacks the verification of GitHub workflow's existence in the repositories. Therefore, the Dangerous-Workflow metric script should be improved to detect empty repositories or repositories without GitHub workflows.

\subsubsection{Pinned-Dependencies}%
In both ecosystems, more than 99\% of packages had a practice of using at least one pinned dependency. Among these, 81\% npm packages and 66\% PyPI packages got a score of 10, indicating that they do not have any unpinned dependencies in the listed directories. 

The score may not reflect an accurate statistic since Scorecard only check \texttt{Dockerfiles, shell scripts}, and \texttt{GitHub workflows} to track dependencies. However, the tool does not check  \texttt{requirements.txt}, \texttt{pyproject.toml}, \texttt{setup.py}, \texttt{package.json}, and  \texttt{package-lock.json} files in PyPI and npm package repositories. %
For PyPI, there are different ways to declare and manage dependencies and their version in Python. For example, \texttt{pyproject.toml} file for declaring dependencies in PyPI is a relatively new standard but not widespread yet. In practice, developers use \texttt{setup.py} (using `setuptools`), which can be non-deterministic and makes it harder to track PyPI dependency. For npm, \texttt{package.json} contains the metadata relevant to the project to manage the project's dependencies, scripts, and versions. %
To depict the accurate status of pinned dependencies in an ecosystem, the Scorecard team should improve the Pinned-Dependencies metrics scripts considering ecosystem standards to evaluate the package dependency.

We also observed that Scorecard does not verify the presence of \texttt{Dockerfiles, shell scripts}, and \texttt{GitHub workflows} files in a repository. If a repository did not have any files of those types, a package would receive a score of 10 for not having an unpinned dependency on those missing files. 
\subsubsection{Vulnerabilities:}%
More than 99\% packages did not have any open vulnerabilities in the OSV database. Hence, they scored 10. %
Scorecard found 7 npm packages and 5 PyPI packages with unfixed vulnerabilities. In addition, 2,703 npm packages and 322 PyPI packages got a score of $-1$ for inconclusive results. Our manual repository review selected repositories where the package had inconclusive scores or open vulnerabilities, ranked by number of dependents. Note that, we did not review packages with scores of 10 since these packages did not have any open vulnerabilities reported in the OSV database. The reason behind the negative score (\texttt{-1}) was that those repositories were empty. In total, we found 39/50 empty repositories. One package had 10 open vulnerabilities with a score of 0, and 9 packages had 1 vulnerability open with a score of 9. %

\subsubsection{Binary-Artifacts} %
More than 99\% packages had a score greater than 0. %
The manual review of 50 repositories found 8 packages with a score of $0$.  The reviewers noticed that these packages had more than 9 binary artifacts with a mean and standard deviation of 78.25 and 87.17, respectively. These packages were umbrella projects encompassing a variety of tools and libraries. Clients are forced to use these binary artifacts directly. %

Another 32 packages in manual review were given a score from $1$ to $10$ based on the number of binary artifacts ranging from 0 to 9. %
A score of 10 means no binaries, a score of 9 means the presence of one binary, and the scores continue to decrease toward 1 as the number of binary artifacts increased toward 9. 
We also found a false positive in one npm package repository, where Scorecard identified 108 binaries, two of which were \texttt{.txt} files. Similar to previous metrics, the Scorecard team should improve the metric script to detect empty repositories along with enhancing the list of binary keywords considering different ecosystems.

\subsubsection{Token-Permissions}%
The metric indicates that npm yielded a more promising result: nearly 84\% of packages have read and write permissions declared in workflows, compared to 71\% of PyPI packages. Our manual review found similar patterns as we observed in Pinned-Dependencies and Dangerous-Workflow. %
Fourteen (14) packages did not have any GitHub Actions specified in the repository, but Scorecard assigned $10$ to those packages for Token-Permissions. Here, the score was \texttt{10} because the tool lacks the verification of GitHub workflow's existence in the repository. The Scorecard team should improve the scripts to detect the presence of GitHub workflows before scoring good or bad practices. 

\subsubsection{License}%
We observed that 68\% of npm packages and 88\% of PyPI packages had published licenses in the GitHub repository, indicating, npm has a higher tendency to avoid licensing in the repository. Our manual review revealed that 4 npm packages and 8 PyPI packages had a license in the repository, specifically in \texttt{Readme.md} and \texttt{setup.py} files. However, Scorecard did not identify them, hence the metric script should be improved to detect licenses more accurately. %

\subsubsection{Code-Review}This check evaluates if the package conducts code reviews prior to merging PR. The first step of the check is to see if Branch-Protection is activated with at least one required reviewer. If this fails, the check looks to see if the last 30 commits are Prow, Gerrit, or Github-approved reviews or if the merge differs from the committer.  30\% of npm packages and 34\% of PyPI packages had code review practices in their repository. A 2022 study~\cite{imtiaz2022phantom} also showed that 52.5\% of the analyzed updates of npm, Crates.io, PyPI, and RubyGems ecosystem packages were only partially code-reviewed, with an overall median code review coverage (CRC) of 27.2\%. One reason behind failing this metric would be that the metric is not applicable if the package has one maintainer. %

However, Our manual review found nine packages scored 0 and had no code review practices even though they had more than one contributor in GitHub repositories. Both ecosystems exhibit a gap in implementing Code-Review in GitHub repositories, indicating that packages contain the risk of introducing non-reviewed code in the software supply chain. %

We also had $-1$ in 5 sample repositories where the repos were empty. To verify this pattern, we reviewed an additional 10 repositories with $-1$. These repositories were empty on GitHub. Hence, indicating why Scorecard assigned $-1$ as an inconclusive result. In total, we found 2,695 (0.3\%) npm and 321 (0.2\%) PyPI empty repositories with $-1$ scoring in Vulnerabilities, Branch-Protection, Packaging, and Signed-Releases and Code-Review metrics.

\subsubsection{Maintained}%
Our findings show that more than 85\% packages in npm and 75\% PyPI packages were unmaintained in GitHub. What is more crucial is that for npm, unmaintained packages may have a more extended period than 90 days, as study ~\cite{zahan2022weak} revealed that in 2021, more than 58\% of packages in the npm registry were unmaintained over two years. Our manual inspections were consistent with Scorecard data where 9/50 packages were inactive in a range of 1 year to 7 years. %

\subsubsection{Branch-Protection}%
Only 10\% of packages passed this metric in each ecosystem, indicating these repos had at least one tier of branch protection applied.  Hence, 90\% npm and PyPI packages had branch protection disabled in the repository. The numbers are considerably high, indicating that a large number of packages in both ecosystems did not create a branch protection rule in repositories. %
Out of five tiers of scoring- ``Enabling branch protection'', ``inhibits force to push, and branch deletion'' are Tier 1 check. Then, the presence of at least one reviewer (Tier 2), enabling status checks (Tier 3), the presence of a second reviewer (Tier 4), and admin dismisses the stale review (Tier 5) are the other tiers. When Scorecard is run without an administrative access token, the requirements that require admin privileges are ignored to avoid penalizing a package score.

Our manual review found that Scorecard metrics only investigate the default branch and any branch that was used for creating a release and uses GraphQL API to verify the protection. However, we verified the branch-protection by looking into the GitHub branches api~\cite{branch_protection}. We found 13/50 packages had a score of $-1$ due to internal error because Scorecard: a) looked for the incorrect branch name that did not exist in the repository;  b) could not locate the branch even though it existed; c) the main branch had a different name than the ``main" or ``master"; and d) branch protections were disabled in main and release branch. %

\subsubsection{Dependency-Update-Tool}%
94\% of npm packages and 97\% of PyPI packages failed this metric because Dependabot and Renovatebot were not used as dependency update tools. A project that uses other tools or manually updates dependencies, will obtain a score of 0 on this metric, just like other packages with outdated dependencies. Dependency-Update-Tool metric call for industry consensus on an ecosystem-wise tool list for Scorecard to report an accurate state.  This metric can only confirm if the dependency update tool is enabled; it cannot confirm if the dependency-update-tool is running or if the tool's pull requests are merged.

\subsubsection{Security-Policy} %
Only 3.2\% npm and 2.5\% PyPI packages have a \texttt{security.md} file. After looking into 50 sample packages, we observed that: a) 25 packages do not adhere to standard security policies; and b) 11 packages have a different reporting procedure for vulnerabilities. Users can, for example, submit bugs in other places such as GitHub issues, specific email addresses, and different bug databases outside of GitHub, or use a different security policy reporting file \texttt{security.rst}. %
Although both ecosystems adopted this practice inadequately, Security-Policy is one of the top recommended GitHub security best practices determined by practitioners~\cite{security_policy} and the SSDF framework also suggested adopting the practice. 

\subsubsection{Packaging}%
Packaging is another metric that indicates the industry consensus is required. Only 1\% of npm packages and 5.8\% of PyPI items passed the packaging metric. Since the software can be packaged in multiple ways, the challenges of coordinating several package release protocols may prohibit developers from releasing packages on GitHub Actions, which can be one reason for the limited number of packaging in the GitHub packaging workflows. At the time of this study, Scorecard did not query the package registries directly. Hence, packages that do not use GitHub Actions get $-1$ instead of $0$. Note that a package's aggregate score will be penalized if it has a score of 0; and inconclusive or $-1$ have no effect on the aggregate score. Our manual inspection identified only 2 npm packages and 6 PyPI packages used GitHub packaging workflow, while 47/50 packages had releases on GitHub. Additionally, the Scorecard failed to detect 4 packages (2 from each ecosystem) that had a publishing GitHub workflow. The names of these files are [publish, ci, release].yml. 

\subsubsection{CII-Best-Practices} %
Scorecard found the CII Best Practices Badge in just 1,665 (0.2\%) npm and 341(0.1\%) PyPI packages. The CII Best Practices program is a way for Free/Libre and Open Source Software (FLOSS) projects to demonstrate that they follow best practices. Projects can voluntarily self-certify to report how they follow each best practice. According to the CII Best Practice Program website %
only 4,766 FLOSS projects have reported their security policies and received different degrees of badges, indicating both ecosystems did not adopt the practice.

\subsubsection{Signed-Releases}%
Similar to the Packaging metric, the Signed-Releases metric's report suggested the need for industry consensus.  Only 578 (0.1\%) npm and 936 (0.5\%) PyPI packages had signed releases. Moreover, almost 100\% packages failed this metric. The low number of Signed-Releases  in GitHub repositories are expected behavior for both ecosystems, as package developers release versions to the package registry (\texttt{npmjs.org or pypi.org}) rather than code hosting platforms like (\texttt{github.com}). Additionally, we observed that GitHub, PyPI, and npm each has different regulations to control package release to a registry. To publish in both registries, the team must take additional steps to confirm the release, which can be incompatible with their workflow~\cite{wermke2022committed}. For instance, the GitHub registry accepts only scoped packages. Therefore, if a JavaScript package is currently named \textit{X}, it must be renamed \textit{@username/X} to publish in GitHub. 

Scorecard assigns $-1$ instead of 0 if the tool can not detect the signed release. In addition, our manual review revealed that Scorecard often verifies older signed versions rather than checking for signatures on the newest five releases. For example, one package received an 8/10 score, meaning 4/5 of recent releases of that package had signed artifacts. However, we found the signed artifacts from older versions, which contradict the defined rules of Scorecard. Then we also observed repositories tagged commits as a release rather than creating a release on GitHub. However, none of the commits were GitHub verified, and Scorecard does not identify tagged releases. Therefore, along with industry consensus, the Signed-Releases metric's scripts need to be updated by Scorecard to improve tool accuracy. %

\begin{figure*}[h]
\centering 
\includegraphics[width=6.0in]{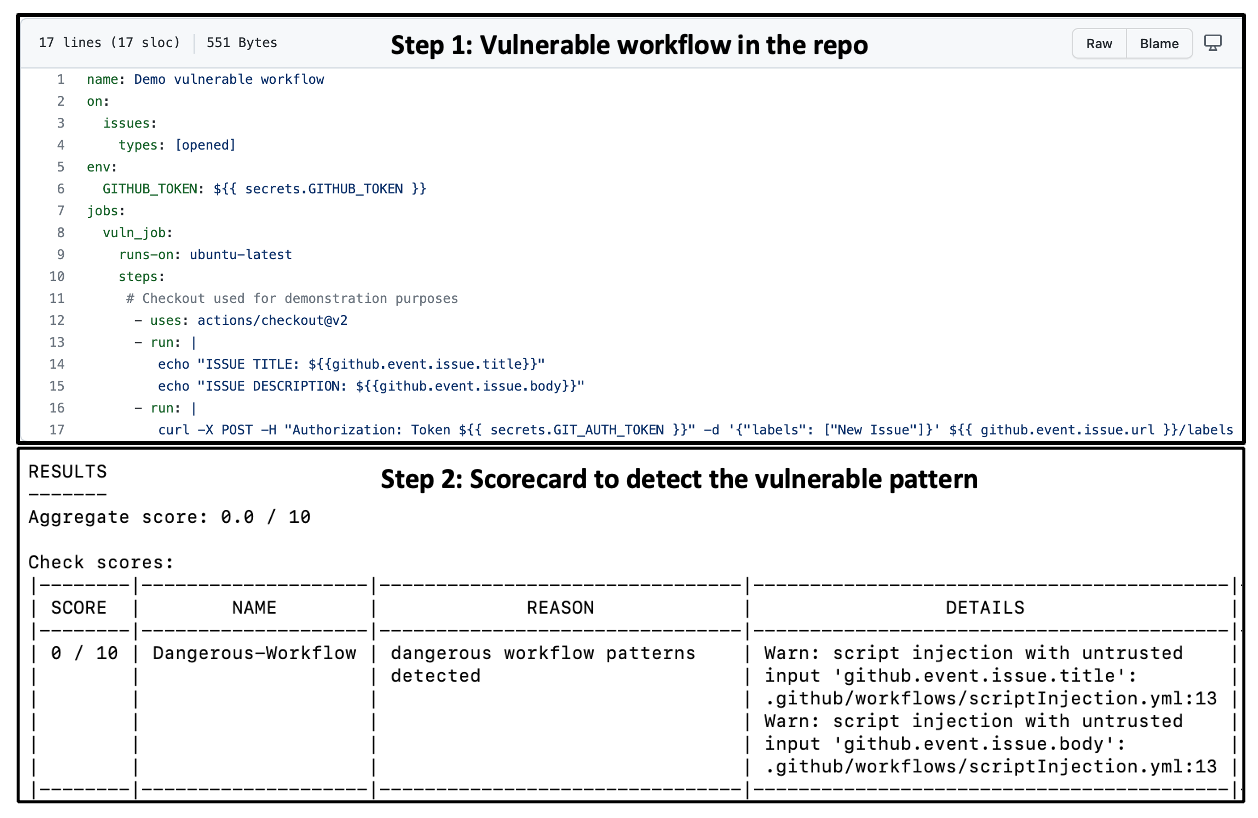} 
\caption{Case study on Dangerous Workflow to detect vulnerable code pattern}
\label{fig:case_study_1_2} 
\end{figure*}

\subsubsection{Fuzzing}Both ecosystems fell short on this metric. Scorecard validates fuzzing exclusively through the tracking of packages in OSS-Fuzz project. OSS-Fuzz has been tested only in 650 open-source packages as of July 2022 %
and a package that uses fuzzing with other tools would fail the check similar to Dependency-Update-Tool metric, indicating why the npm and PyPI ecosystems failed this metric. Out of 650 open-source packages that use OSS-Fuzz, we found 50 npm packages and 104 PyPI packages. Despite the fact that only a few packages passed this metric, PyPI has more fuzzing practice (50 percent more) than npm. One reason why npm packages do not use fuzzing could be that fuzzing JavaScript (JS) engines is tricky and requires expertise. Instead of processing user-supplied seeds, JS engines scan and interpret user seeds into an abstract syntax tree (AST)~\cite{dinh2021favocado} which impacts the performance of fuzzers. Our manual analysis yielded no different results from what we expected. Only two PyPI packages used OSS-Fuzz and 48 packages had no fuzzer. Then, Fuzzing had most of the $-1$ after Signed-Releases and Packaging, but in our manual analysis re-running 12 packages with $-1$, scored $0$ in new run, indicating run time error occurred during the first run. %

\subsection{\textbf{Case study on Dangerous Workflow}}\label{test_case_1} 
In this section, we investigate whether we can exploit GitHub workflows' vulnerable patterns for malicious intent and evaluate whether Scorecard can detect the patterns via the Dangerous Workflow metric. We found 508 PyPI packages (96 packages with an average of 58 dependents) and 1938 npm packages (805 packages with an average of 13 dependents) where packages failed the Dangerous Workflow metric, indicating that their workflows contain vulnerable coding patterns. Even if these repositories are not malicious, potentially dangerous misuse of the workflows may allow malicious attackers access to the data and commit data breaches and theft. %

We refrained from attacking the repositories from an ethical standpoint since many of these repositories are legitimate projects. Therefore, to test whether Scorecard can be used to anticipate malicious attacks, we executed the Scorecard tool on a dummy GitHub repository where we built a workflow with an intentionally-vulnerable \texttt{issue} action, inspired by~\cite{test_case_1}.  %

We execute a reverse shell attack by creating the following \texttt{issue} in the dummy repository from a different GitHub account user- \texttt{New malicious issue title" \&\& bash -i >\& /dev/tcp/4.tcp.ngrok.io/{ngrok endpoint} 0>\&1 \&\& echo"}. 
 Our vulnerable workflow (Figure \ref{fig:case_study_1_2}) in the dummy repository is executed on a GitHub runner whenever a new issue is created by any user. Here, line 14 in step 1 of Figure \ref{fig:case_study_1_2} \texttt{"ISSUE TITLE: \{\{github.event.issue.title\}\}"} is vulnerable to command injection because the hosted runners replace the macros \texttt{\{\{ ... \}\}} blindly and echo \texttt{"\{\{github.event.issue.title\}\}"} becomes echo \texttt{"\{\{New malicious issue title\}\}"}, thus, giving an attacker to run a reverse shell inside the hosted runner as part of the arbitrary code execution capabilities. An attacker can read sensitive files like \texttt{.credential} from the runner folder. %
Step 2 in Figure \ref{fig:case_study_1_2} shows that the Scorecard could identify the vulnerable pattern, referring to the exact line number. The dependents stat of these packages shows that malicious injection may allow attackers to execute supply chain attacks. Therefore, this case study substantiates that the Dangerous Workflow metric is effective for identifying malicious attacks on GitHub workflows.  %

\section{\textbf{Discussion}}
This section discusses the effectiveness and applicability of Scorecard tool security metrics and the gap in npm and PyPI ecosystems security practices. An overview of our discussion is shown in Table \ref{tab:Discussion}.

\begin{table*}[h] \centering \renewcommand{\arraystretch}{1.3}
\caption {Summary of Scorecard Security Metrics Evaluation} \label{tab:Discussion} 
\begin{tabular}{| p{150pt} || p{220pt} |} \hline
\textbf{Scorecard Metrics} & \textbf{Evaluation of this study} \\\hline\hline
Dangerous-Workflow & Effective Metric, Need-for-Improvement  \\\hline
Pinned-Dependencies & Need-for-Improvement \\\hline
Vulnerabilities & Effective Metric \\\hline
Binary-Artifacts & Effective Metric, Need-for-Improvement\\\hline
Token-Permissions & Effective Metric, Need-for-Improvement, Lack of adoption \\\hline
License &   Effective Metric, Need-for-Improvement, Lack of adoption \\\hline
Code-Review &  Effective Metric, Lack of adoption \\\hline
Maintained &   Effective Metric, Lack of adoption \\\hline
Branch-Protection & Effective Metric, Lack of adoption \\\hline
Dependency-Update-Tool & Require Industry consensus \\\hline
Security-Policy &  Effective Metric, Lack of adoption \\\hline
Packaging &   Require Industry consensus \\\hline
CII-Best-Practices &  Require Industry consensus \\\hline
Signed-Releases & Require Industry consensus  \\\hline
Fuzzing &  Require Industry consensus \\\hline
\end{tabular}
\end{table*}

\subsection{\textbf{Effective metrics for both npm and PyPI ecosystems}}
Our study reveals that practitioners can use a subset metrics of Scorecard tool to measure security practices, including metrics- \textbf{Dangerous-Workflow, Vulnerabilities, Binary-Artifacts, Token-Permissions, License, Code-Review, Maintained, Branch-Protection} and \textbf{Security-Policy}. To achieve a higher security score (towards 10), practitioners need to follow the guidelines provided by the Scorecard tool. Most of these are well-established security metrics, also required by the SSDF~\cite{souppaya2022secure} and SCVS~\cite{scvs} framework. However, the guideline provided by Scorecard for Dangerous-Workflow and Token-Permissions are more GitHub-focused, which requires practitioners to implement GitHub workflows to achieve higher scores. Both of these metrics effectively detect security weak links in GitHub workflows, e.g. Dangerous-Workflow can be used to prevent malicious PRs and issues. Therefore, Dangerous-Workflow and Token-Permissions metrics will not be useful to practitioners who do not use GitHub workflows for their CI/CD pipeline.

\subsection{\textbf{Need-for-Improvement metrics}}
 The \textbf{Dangerous-Workflow, Pinned-Dependencies, Binary-Artifacts, Token-Permissions}, and \textbf{License} metrics exhibit the need for Scorecard team's attention for improvement. The Pinned-Dependencies metric requires revision for different ecosystems. For example, Pinned-Dependencies do not check the \texttt{package.json} and \texttt{requirement.txt} and other files for the dependency version, even though the PyPI and npm contain dependency information in such files. License metrics can be improved by enhancing the list of keywords. 
 
Then, accurate ecosystem evaluations require filtering out packages with empty repositories. However, the Scorecard generates aggregate scores for empty repositories. Because Dangerous-Workflow, Binary-Artifacts, Pinned-Dependencies, and Token-Permissions metrics indicate risky patterns in GitHub. For these metrics, empty repositories obtained a score of 10, as repositories were completely devoid of any content, let alone risky patterns. %
Hence, Scorecard assigned a score of $10$ instead of  $0$, or $-1$, whereas the other 11 metrics values were between $0,-1$. Another similar example would be even if a package does not have GitHub workflows, the tool will automatically score $10$ in Dangerous-Workflow and Token-Permissions metrics which do not exactly reflect that the package follows good workflow patterns.  Our findings suggest that Scorecard should check for the existence of GitHub repositories or workflows before reporting on good or bad security practices, since high scores give us a false sense of good security practices. We submitted our findings to the Scorecard team, and the team acknowledged and agreed to improve Scorecard to enable automated testing more effectively in Version 5. 

\subsection{ \textbf{Industry consensus required on metrics}}
 The \textbf{Dependency-Update-Tool, Packaging, CII-Best-Practices, Signed-Releases}, and \textbf{Fuzzing} indicate the requirement of industry consensus before Scorecard can promote these metrics to practitioners.
 Scorecard has proposed guidelines for these practices, but without industry consensus, these metrics hardly have any value from an ecosystem security perspective. Both ecosystems exhibited weak adoption of these practices. 
 
 For example, Scorecard requires practitioners to use specific tools for the Dependency-Update-Tool and Fuzzing metrics to achieve higher scores. However, the industry lacks consensus on the list of tools or research showing the ecosystem's preference regarding those tools. The ecosystem needs to agree on or standardize these tools so that Scorecard can measure the practices. Then, practitioners did not show evidence of using Packaging and Signed-Releases practices on GitHub. The reasons could be that practitioners used the package registry to release the signed/unsigned version and used GitHub as a platform for source code distribution. Either Scorecard could integrate with package registries to collect accurate data on Packaging and Signed-Releases, or practitioners could agree to release signed packages on GitHub, which can be an additional task since different platforms may have different regulations for releasing a package. CII-Best-Practices requires maintainers to self-report their security practices' adherence. Therefore, failing these metrics may not necessarily advocate package owners failed to implement the practices required for CII-Best-Practices, they simply may not have self-reported the practices. %

We do acknowledge, however, that such an industry-wide agreement may be challenging and may take time to implement. %
In that case, Scorecard may separate these metrics from the aggregate score calculation. If a package implements these practices, the package may get bonus points instead of directly impacting the aggregate score. Either way, practitioners and the Scorecard team should address the above-mentioned issues to achieve an accurate picture of ecosystem security practices.

\subsection{\textbf{Ecosystems security comparison}}
In the case of License, Code-Review, and Maintained, PyPI outperformed the npm ecosystem. For example, only 68\% npm packages had a published license in the repository, compared to over 88\% of PyPI packages. %
Although both ecosystems failed the Fuzzing metric check, we found that PyPI exhibited 50\% more fuzzing tools implementation than the npm ecosystem. Then, the Token-Permission metrics showed that npm (84.4\%) has better file permissions in the GitHub workflow compared to the PyPI (71.7\%) ecosystem.

\subsection{\textbf{Lack of adoption in ecosystem-wide security practices}} 
 Both ecosystems indicate lack in practicing \textbf{Token-Permission, License, Code-Review, Maintained, Branch-Protection,} and \textbf{Security-Policy} practices in the GitHub repository. These Scorecard metrics effectively measure security in GitHub (Table \ref{tab:final_data}), but both ecosystems showed inconsistency in adopting these security practices. On the contrary, metrics that require industry consensus demand modification in guidelines proposed by Scorecard, metrics have inheritance reliance on other systems, and more than 90\% packages in both ecosystems scored either 0 or -1.

 Even if the Token-Permission metric needs to be improved, Scorecard identified 15.6\% of npm repositories and 28.2\% of PyPI repositories containing \texttt{yaml} files with write access, indicating package susceptibility to malicious attack. License is important for an organization to comply with organization’s legal policies. 30\% npm packages and 12\% PyPi packages did not contain any valid License in GitHub repositories which is legally require for any organization intending to use those packages. Then both ecosystems lacked to adopt Code-Review (npm: 69\%, PyPI: 65\%) and Maintained (npm: 86\%, PyPI: 76\%) metric, indicating the risk of using unreviewed, unmaintained code. Additionally, around 90\% of the packages in both ecosystems did not show evidence of implementing default Branch-Protection and Security-Policy practices in their repositories.
 
Token-Permission, Code-Review, Maintained, and Security-Policy were all listed by the SSDF framework as important security practices, highlighting the significance of implementing these practices.

\section{\textbf{Limitations and future work}}
In our study, we group security scores into three categories ([-1,0,1-10]) to avoid arbitrary scoring bias; some metrics scoring may not be representative of the severity of security risk. For example, the vulnerabilities metric looked for open vulnerabilities in OSV database and assigned scores based on number of open vulnerabilities. However, the tool does not look into the severity of vulnerabilities. If a package has one exploitable vulnerability it will score 9, whereas a package with more than 9 open but non-exploitable vulnerabilities will score 0. Even though 9 seems like a better score, severity-wise, it is a high-risk package with exploitable open vulnerabilities. Additionally, if a package did not contain any vulnerabilities reported in OSV database, the package will receive a score of 10, which does not confirm that the package is free of vulnerabilities. Although this scoring is a limitation of the Scorecard tool, and we tried to reduce the bias by grouping the score into three categories, metrics like vulnerabilities may prevent us from achieving accurate security status of these ecosystems.

Then, our case study on Dangerous Workflow on a dummy repository may not represent 2,446 packages, which is a limitation of this study. One future direction of this research is to confirm whether all the repositories that failed in Dangerous Workflow metrics are vulnerable to malicious attacks. While our case study shows that vulnerable patterns identified by Scorecard are exploitable, without verifying each repository and their GitHub action individually, it is hard to confirm whether all repositories are prone to malicious attacks.

Additionally, our result may not be representative in the future since Scorecard is evolving and practitioners are following Scorecard guidelines. Therefore, the results are subject to change. However, the finding will assist the Scorecard team, ecosystems, and practitioners in improving their current state. Another future direction of our study is to reach out to npm and PyPI practitioners to validate whether they have decided not to follow security practices consciously, use alternative practices, or have other challenges preventing them from adopting security practices.

\section{\textbf{Conclusion}}This study compares the npm and PyPI ecosystems' security practices in GitHub repositories using Scorecard tools. Our work focuses on measuring and understanding the adoption of cross-ecosystem package security practices. We also evaluated whether we can leverage the Scorecard tool metrics to measure ecosystem-wide automated security practices. We found that 13 Scorecard security metrics were compatible with the SSDF framework. Next, we identified 9 Scorecard security metrics that can be used to measure npm and PyPI package security.  Then, ``Dangerous Workflow'' can aid in identifying malicious attacks. Five practices, however, necessitate industry agreement. Both ecosystems showed gaps in implementing Token-Permission, License, Code-Review, Maintained, Branch Protection, and Security Policy practices. %

Knowing about these security practices and their challenges will inspire and direct practitioners on what to do to adopt these practices or identify the gaps preventing them from doing so. We have also observed and been told that the Scorecard team welcomes new security metrics and discussions that indicate the Scorecard is evolving with time. Therefore, our study aims to draw practitioners' attention to creating action plans to enhance Scorecard security metrics for assessing ecosystem-wide security practices. To improve the tool's ability to measure security automatically, the ecosystem managers, the Scorecard team, and the practitioners can drive ecosystem-wide standards. Such industry-wide consensus will push software producers to start implementing those practices.

\section{ACKNOWLEDGMENTS}
This work was funded by Cisco and National Science Foundation Grant No. 2207008. Any opinions expressed in this material are those of the author(s) and do not necessarily reflect the views of the National Science Foundation. We thank the OpenSSF Scorecard Team for their valuable feedback and assistance in generating Scorecard data for such a vast number of repositories.

\bibliographystyle{IEEEtran}
\bibliography{reference}

\end{document}